# Adaptive Q-control for Tapping-mode Nano Scanning Using a Piezo-actuated Bimorph Probe

**Ihsan Gunev, Aydin Varol, Sertac Karaman, Cagatay Basdogan**

College of Engineering, Koc University, Istanbul, Turkey, 34450.

**Abstract:** A new approach, called *Adaptive Q-control*, for tapping-mode Atomic Force Microscopy (AFM) is introduced and implemented on a home-made AFM set-up utilizing a Laser Doppler Vibrometer (LDV) and a piezo-actuated bimorph probe. In the standard Q-control, the effective Q-factor of the scanning probe is adjusted prior to the scanning depending on the application. However, there is a trade-off in setting the effective Q-factor of an AFM probe. The Q-factor is either increased to reduce the tapping forces or decreased to increase the maximum achievable scan speed. Realizing these two benefits simultaneously using the standard Q-control is not possible. In adaptive Q-control, the Q-factor of the probe is set to an initial value as in standard Q-control, but then modified on the fly during scanning when necessary to achieve this goal. In this paper, we present the basic theory behind the adaptive Q-control, the electronics enabling the on-line modification of the probe's effective Q-factor, and the results of the experiments comparing three different methods: scanning a) without Q-control, b) with the standard Q-control, and c) with the adaptive Q-control. The results show that the performance of the adaptive Q-control is superior to the other two methods.

# 1. Introduction

Atomic Force Microscopy (AFM), introduced by Binning et al.[1], is a widely used imaging and manipulation technique at both micro and nano-scale. In AFM, a micro fabricated cantilever-shaped probe with a sharp tip underneath is used to scan substrate surface. The combination of attractive and repulsive interaction forces between the tip and the substrate causes the free end of the probe to deflect. Through a measurement feedback and positioning of the sample along the vertical direction perpendicular to the substrate surface, 3D topographic map of the surface can be obtained in resolutions below nano-meters. There are three modes of operation in AFM: contact, non-contact, and tapping modes (see the review in Ref. 2).

The tapping-mode AFM, first introduced by Zhong et al.[3], has been investigated extensively in earlier studies[4, 5]. In tapping-mode AFM, a probe oscillating at free-air amplitude ($A_0$) near to its resonant frequency ($\omega_R$) is scanned over the sample surface. As the name "tapping mode" suggests, the tip of the cantilever taps the sample surface for a very short period of time reducing the oscillation amplitude to $A<A_0$ [3]. In amplitude modulation scheme, a scan controller moves the sample or the probe in vertical direction (i.e. Z-direction) such that the tapping oscillation amplitude A stays constant at a pre-set amplitude $A_{set}$. In frequency modulation scheme, the shift in the resonance frequency is measured and used in the feedback loop.

One of the benefits of tapping-mode AFM is that the probe applies small lateral forces to the sample and hence damaging the sample in the lateral direction is prevented[6]. However, tapping forces may reach to high amplitudes and easily cause damage on the sample and the probe tip. This is more critical when scanning soft samples such as the ones used in biological applications[6, 7], but the above argument applies to all kinds of samples. In order to limit the tapping forces, one approach is to keep the $A_{set}$ close to $A_0$[7]. However, in that case, the probe detaches from the surface easily and error saturation becomes a major problem limiting the scan performance. For example, when the probe encounters a steep downward step during scanning, it suddenly detaches from the sample surface and the magnitude of the error between the oscillation and set amplitudes saturates at ($A_{set} - A_0$), which limits the speed of the controller response in Z-direction (see the detailed discussion of error saturation problem in Ref. 8) Therefore, it takes longer time for the tip oscillations to settle down $A_{set}$ again. As a result, the resulting scan profile during the saturation period is erroneous. Another approach for reducing the tapping forces is to use probes having high Q-factor. The Q-factor of a probe is inversely proportional to the amount of damping in the system and a measure of its energy dissipation in each oscillation cycle. The relation between the variables discussed above is given as [7, 9, 10],

$$\langle F_{ts} \rangle \propto \frac{k}{Q} \sqrt{\left(A_0^2 - A_{set}^2\right)} \qquad (1)$$

where, $F_{ts}$ represents the tapping forces, Q is the effective Q-factor, and k is the stiffness of the probe. Using a probe with high Q-factor has shown to reduce the magnitude of the tapping forces[11]. However, AFM probes having high Q-factor dissipate less energy at each oscillation cycle; therefore their response is slower than the probes having low Q-factor.

In order to have a probe with a high Q-factor, either the physical dimensions of the probe is reduced or the probe is operated in vacuum. However, these approaches are not always practical. For this reason, Q-control is introduced in order to change the effective Q-factor of a probe actively[12]. In Q-control, the effective Q-factor of a cantilever probe is set to a desired value prior to scanning by means of an extra electronic circuit, including a phase shifter and a gain amplifier, in the feedback loop. Using this circuit, the measured displacement signal of the probe is phase shifted and added to the actuation signal after it is scaled by a constant gain. A detailed theoretical analysis of Q-control and its effect on cantilever dynamics are provided by Rodriguez and Garcia[7]. These theoretical results are extended by Holscher et al.[13] to include the tip-sample interactions based on nonlinear effects. The effect of controlling Q-factor on the sensitivity of oscillation amplitude and phase signal is investigated by Kokavecz et al[14]. Increasing the sensitivity of the probe using a phase shifter without affecting the oscillation amplitude is suggested by Tamayo and Lechuga[15]. The imaging bandwidth of a tapping-mode AFM probe under Q-control is analyzed by Kokavecz et al[16]. Sulchek et al.[8, 12] showed that the sensing bandwidth of a probe and scan speed can be improved significantly by actively lowering the Q-factor of the probe. However, increased energy dissipation at each oscillation cycle causes the tip to tap harder onto the sample when the Q-factor is lower. In working liquid, adjusting the Q-factor of the probe is also beneficial[17, 18, 19]. Due to the high viscous damping of the liquid, the effective Q-factor of the probe is reduced naturally when it is operated in liquid. However, this may cause the probe tap harder on the sample. In order to prevent the probe tip damaging the soft sample, the effective Q-factor of the probe is increased using the standard Q-control approach.

In all of these studies involving Q-control, the Q-factor of the probe is adjusted prior to scanning such that either the maximum achievable scan speed is increased or the tapping forces are reduced. Achieving these two benefits simultaneously is not possible using the standard Q-control due to the trade-off between the bandwidth of the probe and the magnitude of the tapping forces. This paper proposes a new method, called *adaptive Q-control*, for adjusting the Q-factor of a piezo-actuated bimorph probe dynamically during scanning when necessary. In our approach, Q-factor of the probe is modified adaptively on the fly depending on the profile of the surface being scanned. In this paper, we show that adaptive Q-control solves the error saturation problem and allows us to achieve higher scanning speeds with lower tapping forces than that of the standard Q-control. As an alternative solution to the same problem, a variable PID controller is suggested by Kodera et al[20]. In this approach, the error between A and $A_{set}$ is amplified artificially by multiplying it with a constant gain when the oscillation amplitude A is higher than a threshold amplitude, $A_{threshold}$. This amplified error signal is then fed back to the scan controller to increase the magnitude of the control signal. As a result, the time elapsed during error saturation is shortened, the detachment of the probe from the surface is prevented, and the maximum achievable scan speed is increased. In our approach, the Q-factor of the probe is modified adaptively on the fly using a new feedback circuit when the oscillation amplitude A exceeds $A_{threshold}$. An increase in Q-factor results in an increase in free-air amplitude $A_0$ which in turn prevents the error saturation and accelerates the controller response.

We have developed an AFM set-up for scanning nano-scale surfaces and an analog signal processing unit for adjusting the Q-factor of the piezo-actuated bimorph probe dynamically. We then compared the scanning performances of 1) conventional PI scan-controller without Q-control, 2) PI scan-controller with standard Q-control, and 3) PI scan-controller with the proposed adaptive Q-control. The remainder of the paper is organized as follows: In Section 2, we discuss the fundamental principles of standard Q-control and introduce the concept of adaptive Q-control. In Section 3, we discuss the components of our home-built AFM set-up and the electronics enabling the on-line modification of Q-factor of the probe. We investigate the dynamical characteristics of the piezo-actuated probe in Section 4. In Section 5, we compare the results of the scanning experiments performed using the three different control schemes mentioned above. Finally, the discussion of the results and the conclusions are given in Section 6.

## 2. Adaptive Q-Control

If an AFM probe is modeled as a spring-mass-damper system, its equation of motion can be written as

$$\left.\begin{array}{c} m\ddot{z} + b\dot{z} + kz = F_{ts} + F_o \cos(\omega_R t), \\ \omega_n = \sqrt{\dfrac{k}{m}} \\ Q = \dfrac{m\omega_n}{b} \\ m\ddot{z} + \dfrac{m\omega_n}{Q}\dot{z} + kz = F_{ts} + F_o \cos(\omega_R t) \end{array}\right\} \quad (2)$$

where, z is the vertical displacement of the probe, m, b, k, and Q are the effective mass, damping coefficient, spring constant, and quality factor of the probe for the $n^{th}$ vibration mode respectively, $\omega_n$ is the natural frequency, $\omega_R$ is the corresponding resonance frequency, and $F_0$ is amplitude of the external driving force. If there is no tip-surface interaction ($F_{ts} = 0$), Eq. 2 is the set of equations of a forced harmonic oscillator and the resonance frequency $\omega_R$ is approximately equal to $\omega_n$ for high values of Q-factor (note that $\omega_R = \omega_n \sqrt{1 - 1/2Q^2}$). The main idea behind the standard Q-control is to modify the quality factor (i.e. effective damping) of the probe by adding a feedback signal to the actuation signal as given in Eq. 3,

$$\left.\begin{array}{c} m\ddot{z} + b\dot{z} + kz = F_{ts} + F_o \cos(\omega_R t) + G\dot{z} + Hz \\ m\ddot{z} + (b-G)\dot{z} + (k-H)z = F_{ts} + F_o \cos(\omega_R t) \\ m\ddot{z} + \dfrac{\omega_n^*}{Q^*}\dot{z} + k^*z = F_{ts} + F_o \cos(\omega_R t) \\ k^* = (k-H) \\ \omega_n^* = \sqrt{\dfrac{(k-H)}{m}} \\ Q^* = \dfrac{m\omega_n^*}{(b-G)} \end{array}\right\} \quad (3)$$

where, G and H determine the amount of velocity and position feedback gain respectively. Eq. 3 shows that pure velocity feedback (H = 0) only modifies the Q-factor of the system whereas any position feedback leads to the modification of the Q-factor as well as the natural frequency $\omega_n$. In commercial AFM systems, vertical oscillations of the probe are measured using a photo detector and then a phase-shifter is employed to shift the displacement signal by $\pi/2$. Finally, the phase-shifted signal is multiplied by the constant gain G and added to the actuation signal to modify the Q-factor of the probe (see Figure 1).

Depending on the application, the value of G is set prior to scanning to either achieve the desired scan speeds or the desired tapping forces since there is a tradeoff between them. Achieving higher scan speeds requires low Q-factor which causes excessive tapping forces while light tapping with high Q-factor slows down the scanning process. Applying low tapping forces to the sample under high scan speeds is desirable but this leads to the error saturation problem discussed earlier. To overcome this problem, we propose to use the adaptive Q-control. In adaptive Q-control, the Q-factor of the probe is adjusted on the fly during scanning when necessary. In order to reduce the free air saturation problem that occurs when the probe encounters a downward step during scanning, the adaptive Q-controller increases the Q-factor of the probe by adjusting the gain G in real-time as

$$\left. \begin{array}{ll} G(t) = G_0(A - A_{threshold}) + G_{set}, & \text{if } A > A_{threshold} \\ G(t) = G_{set}, & \text{else} \end{array} \right\} \quad (4)$$

where, $G_0$ is a constant gain and $G_{set}$ is the gain value corresponding to the desired Q-factor of the probe $Q_{set}$ (i.e. $Q_{set}$ is the effective Q-factor of the probe as in standard Q-control). If $G_{set} = 0$ and when $A < A_{threshold}$, the Q-factor of the probe is equal to its native value at the resonance frequency $\omega_R$. When the oscillation amplitude A exceeds a certain amplitude threshold $A_{threshold} > A_{set}$ and tends to reach the free-air amplitude during scanning of downward steps, the velocity signal is first multiplied by the gain G(t) and then added to the actuation signal to prevent the error saturation problem. As a result, the effective Q-factor of the probe increases momentarily which in turn increases the oscillation amplitude of the probe. This causes a sudden increase in the magnitude of the error signal ($A_{set}$ - A) and the control signal sent to the Z-actuator adjusting the position of the sample with respect to the probe tip. The system responds faster to the downward steps and the error saturation problem is significantly reduced, leading to scan speeds faster than the conventional PI controller with or without the standard Q-control.

**3. Set-up**

We have developed a new AFM set-up operating in tapping mode to implement the proposed controller. The components of the set-up are shown in Figure 2. All movements of the sample relative to the scanning probe is controlled using a XYZ nano-stage (PI Inc, Germany, Model No: P-517.3CD) combined with a 3 channel digital controller unit (E-710.P3D). The nano-stage has a travel range of 100 x 100 x 20 μm and equipped with integrated capacitive sensors for precise positioning. The digital controller of the stage is connected to a digital data acquisition (DAQ) card (PCI-DIO-96, National Instruments Inc.) via parallel input/output port running a servo loop at 5ms/cycle. The usage of a digital nano-stage brings high accuracy positioning, noiseless operation, and accurate measurement of XYZ position. Moreover, it significantly reduces the non-linearity and creep typically observed in open-loop scanning systems utilizing a piezo tube (PZT) as the positioning device.

A commercially available self-actuated AFM probe (DMASP, Veeco Probes Inc., Santa Barbara, CA) is used for scanning. The probe contains a ZnO stack (consisting of 0.25μm Ti/Au, 3.5µm ZnO, and 0.25µm Ti/Au) at its base[21]. This stack, along with the silicon cantilever, acts as a bimorph actuator to oscillate the probe up and down (Figure 3(a)). When voltage signals are applied to the pads at the fixed end of the cantilever, the ZnO layer expands or contracts according to the piezoelectric phenomena causing the bimorph probe tip to oscillate. In this way, the cantilever can be resonated by applying voltage signals at a desired frequency. The probe which is already glued and wire bonded on a chip by the manufacturer (Figure 3(b)) is mounted on a manual XYZ stage (462 Series, Newport Inc., Germany) to bring it sufficiently close to the sample surface.

An electrical signal for the actuation of the probe is provided by a signal generator (Agilent Technologies Inc, Model No: 33220A). A Laser Doppler Vibrometer, LDV, (Polytec GmbH, Germany.) is used for measuring the vertical vibrations of the probe. The LDV having a bandwidth of 1.5GHz, measures the out of plane velocity of a point on the probe by collecting and processing the back-scattered laser light. It is composed of a controller (OFV-5000) and a fiber interferometer (OFV-551). The sensor head of OFV-551 delivers a He-Ne red laser beam ($\lambda = 633$ nm) to a measurement point on the probe and collects the reflected light. Using an optical microscope (VM-1V, Meiji Techno Co., Ltd, Japan) with 5X lens, the laser beam is focused down to a spot size of ~2 µm. The controller OFV-5000 processes the data collected from OFV-551 using a wide bandwidth velocity decoder (VD-02) having a resolution of 0.15 µm/s. The measurement data is acquired from the LDV controller (OFV-5000) in scaleable units of mm/s/V. In addition, Polytec OFV-71 and OFV-72 units are mounted between the microscope and the CCD camera (Flea, Point Grey Research Inc., Vancouver, Canada) to integrate the microscope system with the OFV-551. OFV-71 contains movable mirrors to deflect the laser beam so that one can manually position the laser spot in the area of view (AOV) of the microscope.

An analog signal processing circuit (Figure 4) consisting of a) a root mean square (RMS) converter, b) a variable phase-shifter, and c) a voltage-multiplier is built and integrated into the AFM system to adaptively modify the Q-factor of the probe on the fly during scanning.

*a) RMS converter*: In order to obtain the vibration amplitude of the probe from the measured velocity signal in real-time, an analog RMS converter is used. Since, an AFM probe is typically vibrated at or close to its resonant frequencies for better scanning results, the acquisition of complete sinusoidal velocity signal requires high-speed sampling. This requirement is eliminated by using an integrated circuit (AD536AJH, Analog Devices, Norwood, MA) which computes the RMS value of the vibrometer output at update rates higher than the sensing bandwidth of the probe (Figure 4). The capacitor C2 shown in Figure 4 adjusts the width of the running RMS window. Choosing a low value for C2 improves the bandwidth of the RMS converter but increases the ripple in the output. The capacitor C3 acts like a single-pole post-filter and increases the quality of the output signal and the bandwidth of the converter.

*b) Phase-shifter*: An analog phase-shifter is integrated into the signal processing circuit to eliminate the intrinsic time delay in the LDV and hence to obtain a true velocity signal. Although the LDV used in our set-up is equipped with a velocity decoder and directly measures the vibration velocity of the probe, we observed a constant time delay (~4μs) in the measurement signal, which causes an additional phase lag between the actuation and output signals. Further experiments and the discussions with the manufacturer of the LDV confirmed that the source of this delay is the time spent by the LDV controller for the digital signal processing. To eliminate the phase lag due to this constant time delay, the resistance in the phase shifter ($R_{var}$ in Figure 4) is adjusted in advance such that the output comes from the phase shifter is always in phase with the true velocity signal.

*c) Voltage-multiplier*: In order to multiply the phase-corrected velocity signal by the variable gain G(t) as discussed in Section 2, an analog voltage multiplier (AD633, Analog Devices, Norwood, MA) is integrated into the signal processing circuit. This unit enables us to modify the Q-factor of the probe on the fly during scanning when necessary. The transfer function of the AD633 chip, supplied by the manufacturer, is W(t)=[V(t)*G(t)]/10+P(t), where in our case, P(t) is the periodic actuation signal coming from the signal generator at or close to the resonant frequency of the probe (i.e. $F_0\cos(\omega_R t)$ in Eq. 2), V(t) is the phase-corrected velocity signal (i.e. $\dot{z}(t)$ in Figure 1), G(t) is the variable feedback gain adjusted by the adaptive Q-controller running on an xPC Target (The MathWorks, Inc.), and W(t) is the actuation signal sent to the probe. Using AD633, the measurement signal is first multiplied by G(t) and then added to the actuation signal to modify the Q-factor of the probe.

The xPC Target runs a real-time controller which takes the RMS of the velocity signal as input and outputs the feedback gain G(t). The software for the controller is developed in Matlab/Simulink (The MathWorks, Inc.) and then converted to C code using the Real-Time Workshop (The MathWorks, Inc.) and compiled by Microsoft Visual Studio C++. The computer running the xPC Target is a high-performance PC equipped with a DAQ card (PCI-6025E, National Instruments Inc.) which enables the data communication between the xPC Target and the signal processing circuit.

In order to scan a sample using the developed AFM system, the velocity of the probe tip is measured as a continuous signal using the LDV and then this signal is converted to an RMS signal and sampled by a DAQ card (PCI-6034E, National Instruments Inc.) into a computer. The oscillation amplitude of the probe A is calculated from the RMS signal and then compared with the desired oscillation amplitude $A_{set}$. A scan-controller (PI-controller) developed in LabVIEW (National Instruments Inc.) keeps the vibration amplitude of the probe constant during raster scanning by moving the nano stage up and down along the Z-axis based on the error signal ($A_{set}$ - A).

## 4. Dynamical Characterization of the Probe

The native Q-factor of the probe for the first three vibration modes is calculated from the amplitude versus frequency response curves by measuring the full width at half maximum (FWHM) bandwidth. The Q-factor of the probe at the resonance frequency $\omega_R$ is calculated from the ratio of $\omega_R$/FWHM (see Table I). Alternatively, the Q-factor of the probe can be calculated using the amplitude modulation method[9]. The mode shape of the probe at the second resonance frequency is torsional; hence it cannot be used for scanning in tapping-mode AFM. The first and third modes are flexural and the third one is preferred over the first one due to its smaller time constant $T = Q/\omega_R$, where $1/T$ is defined as the sensing bandwidth of the probe. Cantilevers with higher bandwidth responds faster to the perturbations and enable scanning at higher speeds. As it can be seen from the Table I, the bandwidth of the third mode is higher than that of the first mode.

In order to observe the dynamical response of the probe at the third resonance mode ($\omega_R$ = 210 kHz) for different feedback gains G, the Q-factor of the probe is modified using the signal processing circuit and then measured experimentally (Figure 5). It is observed that the effective Q-factor of the probe changes exponentially with the gain.

## 5. Scanning Experiments

A calibration grating having periodic steps with a height of 104±2 nm (TGZ Series, Mikromash, USA) is scanned using the set-up at different scan speeds and set amplitudes. The same profile is scanned using three different methods: 1) PI controller without Q-control, 2) PI controller with standard Q-control, and 3) PI controller with the proposed adaptive Q-control.

Figure 6 shows the profile and the error signal ($A_{set}$- A) of a single scan line for the scan speeds of X = 2 μm/s and 5X respectively when the standard PI controller is employed for scanning without Q-control. Due to the error saturation, the response of the PI controller at high speeds is slow and causes an inclined profile for the downward steps (see Figure 6(b)). The adverse effects of the error-saturation on the scan profile amplify as the scan speed is further increased. This limits the maximum achievable scan speed.

The adaptive Q-controller overcomes the error saturation problem. It automatically increases the effective Q-factor of the probe during scanning when the error between the oscillation amplitude A and the set amplitude $A_{set}$ is close to the saturation limit, but higher than the threshold (i.e. A > $A_{threshold}$). Augmenting the effective Q-factor of the probe increases its oscillation amplitude and the magnitude of the error signal sent to the standard PI controller to adjust the Z-movements of the sample with respect to the probe tip. As a result, the scanning system responds to the downward steps faster. This improves the overall quality of the scan profile and makes it possible to scan faster without sacrificing the image quality. Figure 7 shows the profile and the error signal of a single scan line for the scan speed of 5X when the adaptive Q-control is active. The undesired effects of the error saturation on the profile are significantly reduced with the adaptive Q-control.

Figure 8 shows the effect of turning the adaptive Q-control on. The solid line represents the RMS of the vibrometer output whereas dashed line shows the value of the feedback gain G(t). Figure 8(a) pinpoints the regions of error saturation where the error cannot exceed the maximum saturation amplitude of ($A_{set}$-$A_0$). When the adaptive Q-controller is active as shown in Figure 8(b), the G(t) is increased automatically (and hence the Q-factor of the probe) if the error saturation is detected. An increase in the Q-factor causes an increase in the vibration amplitude A (as well as the magnitude of the error signal $A_{set}-A$) and results in faster response of the controller.

An alternative way to scan a sample at high speeds is to reduce the native Q-factor of the probe prior to scanning using the standard Q-control. However, reducing the Q-factor of the probe amplifies the tapping forces applied by the probe tip to the sample (see Eq. 1). In order to reduce the tapping forces, the ratio of $A_{set}/A_0$ can be increased to a value close to unity (see Eq. 1), but this leads to the error saturation problem and easy detachment of the probe tip from the sample surface. In addition to increasing the scanning speed, the adaptive Q-controller reduces the tapping forces by enabling us to use higher set amplitudes ($A_{set}$) for scanning. In Figures 9(a) and 9(b), we present the profile of steps scanned using the standard and the adaptive Q-control respectively. In both cases, the effective Q-factor of the probe is reduced to the same value to achieve higher scan speeds and the set amplitude ($A_{set}$) is chosen close to the free air amplitude $A_0$ to reduce the tapping forces. As shown in the Figure 9, the adaptive Q-control outperforms the standard Q-control.

## 6. Discussion and Conclusion

In this paper, the adaptive Q-control is proposed and implemented. In order to test the performance of the proposed controller, an AFM set-up utilizing a LDV and a piezo-actuated bimorph probe is constructed. The scanning experiments using the AFM set-up demonstrate that the adaptive Q-control increases the maximum achievable scan speed of the system while keeping the magnitude of the tapping forces low. Achieving these two benefits simultaneously using the standard Q-control is not possible due to the trade-off in adjusting the Q-factor of an AFM probe. In order to scan a sample faster without sacrificing the image quality, the Q-factor of the probe should be reduced prior to scanning which may cause the probe to apply excessive tapping forces to the sample. In order to tap the sample lightly, either the Q-factor of the probe should be increased which increases the sensitivity of the probe but reduces the scan speed or the ratio of $A_{set}/A_0$ should be increased which causes the error saturation problem. For the scenario given in Figure 9, the Q-factor of the probe is reduced to increase the scan speed ten times of the initial operating speed and also the ratio of $A_{set}/A_0$ is increased to reduce the tapping forces simultaneously. When a downward step is encountered and the set amplitude $A_{set}$ is close to the free air amplitude $A_0$, the error signal is weak and the response of the scan controller (i.e. PI controller) is slow. Moreover, the sensitivity of the probe is reduced and hence it cannot quickly respond to the changes. The resulting scan at these extreme settings is highly unsatisfactory even if the standard Q-control is used (Figure 9(a)), but much better when the adaptive Q-control is turned on (Figure 9(b)).

The technologies that enable us to modify the Q-factor of the probe in real-time are the piezo-actuated scanning probe used in our AFM set-up and the analog signal processing circuit designed to work with it. The piezo actuation layer on the probe is much smaller in size and mass than a piezo stack typically used in standard AFM systems; therefore its response is order of magnitude faster[22]. The wide actuation bandwidth of piezo layer enables us to rapidly modify the effective Q-factor of the probe on the fly during the scanning when necessary. In addition, the new signal processing circuit adjusts the variable gain G(t) automatically, which is controlled by an xPC Target for achieving real-time performance. In comparison, the gain in the standard Q-control is set by the user prior to scanning and does not change throughout the scanning process. In adaptive Q-control, this gain is adjusted by the controller on the fly depending on the error in amplitude signal. If there is a tendency towards saturation in error, the controller increases the Q-factor of the probe rapidly to avoid the problem.

We should also emphasize that the LDV in our set-up is equipped with a velocity decoder and hence enables us to feed the "true" vibration velocity of the probe to the actuation signal in order to modify the effective damping of the probe. On the other hand, in a typical AFM system utilizing a standard Q-controller, displacement of an oscillating probe is measured using a photo-detector first, shifted in phase, then scaled by a gain, and finally used as the velocity signal in the feedback loop. This approach works well if the oscillation profile of the probe stays sinusoidal during scanning. Since the interactions between the probe tip and substrate surface are nonlinear during tapping and the oscillations are not perfectly sinusoidal, the LDV equipped with a velocity decoder provides a more realistic velocity feedback than the conventional phase-shifting approach.

**Tables**

TABLE I. The dynamical characteristics of the probe for the first three resonance modes.

| Mode | Mode Shape | Resonance Frequency (Hz) | Quality Factor | Bandwidth (Hz) |
|---|---|---|---|---|
| 1st | 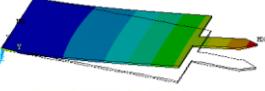 | 48.600 | 129 | 376 |
| 2nd | 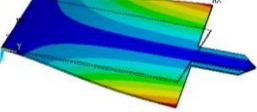 | 180.000 | - | - |
| 3rd | 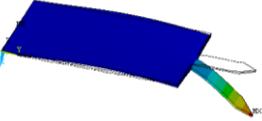 | 210.000 | 311 | 675 |

**Figures**

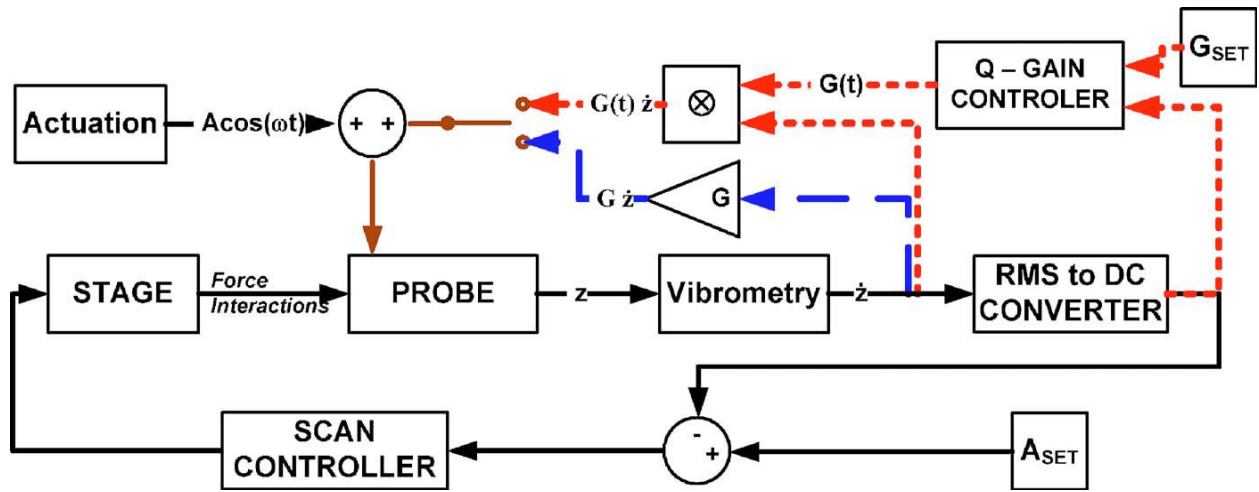

FIG. 1. Block diagram of the nano-scanning system. The dashed lines show the standard Q-control with constant feedback gain G, whereas dotted lines show the adaptive Q-control with variable feedback gain G(t).

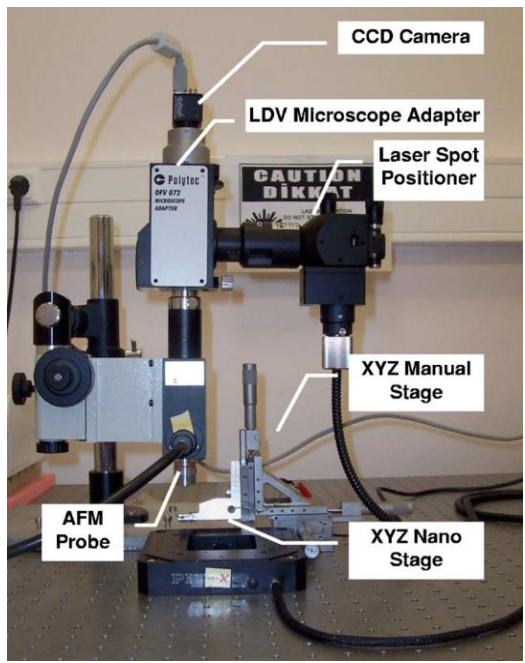

FIG. 2. The components of the developed AFM set-up.

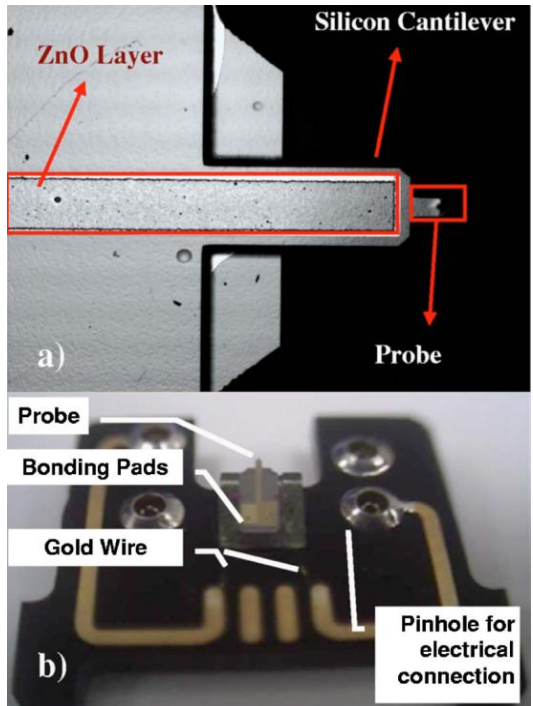

FIG. 3. a) The image of the probe used in the set-up under optical microscope (tip side is up). b) The probe is supplied by the manufacturer as glued and wire bonded on a chip.

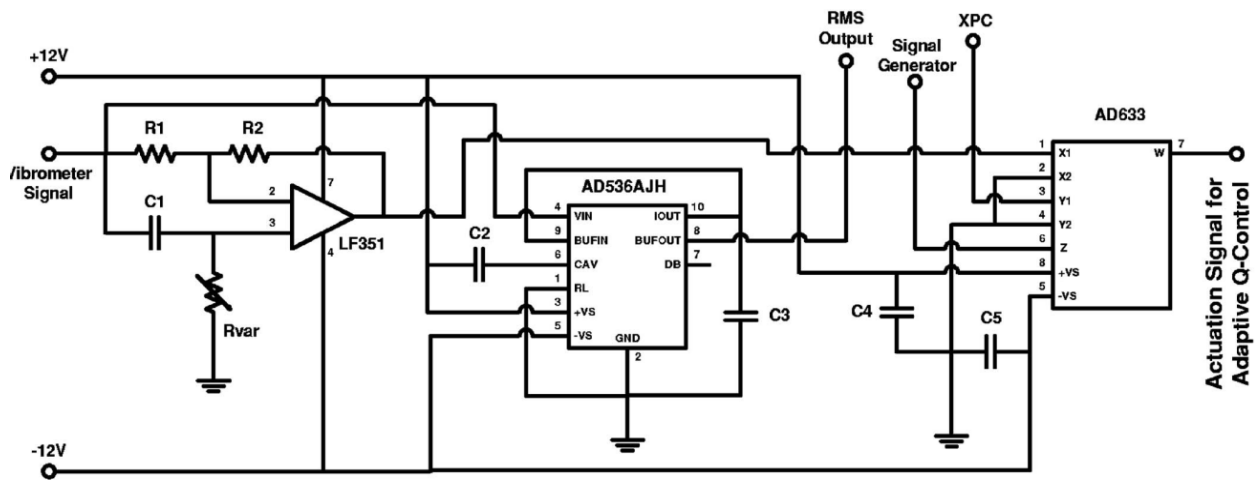

FIG. 4. The signal processing circuit.

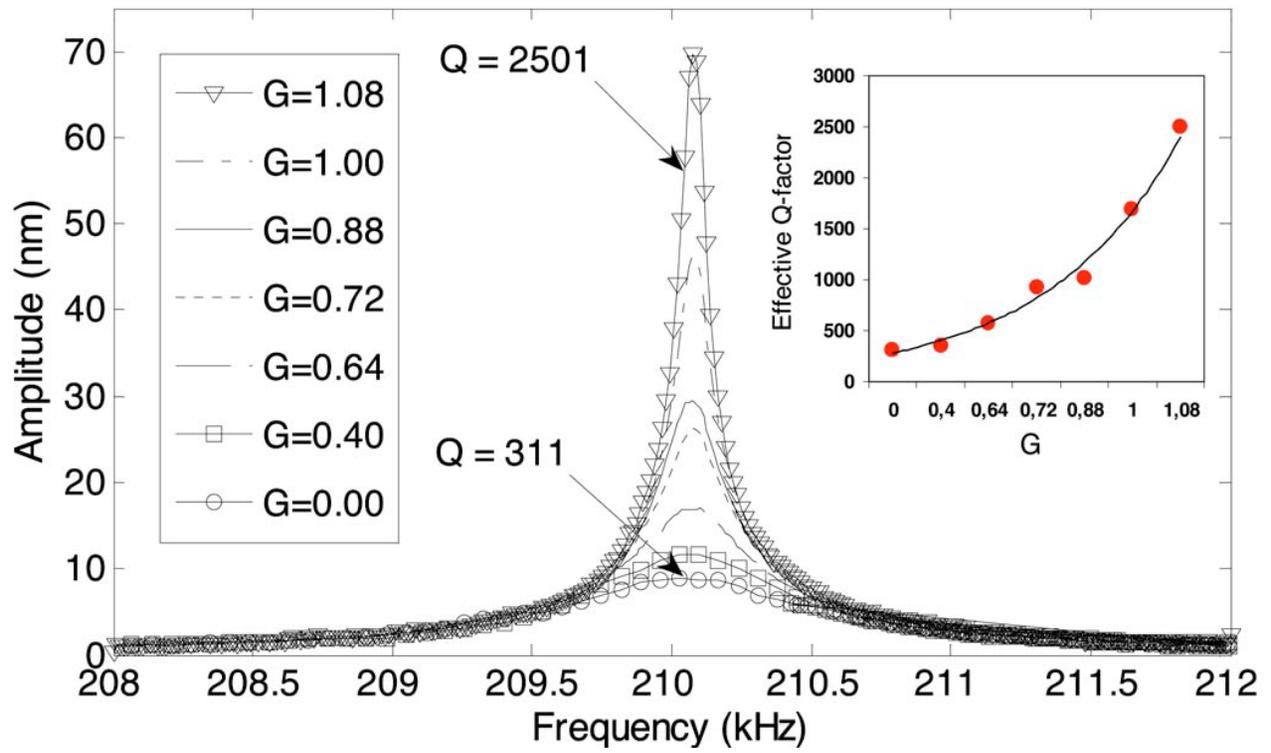

FIG. 5. The frequency sweep curves for various values of feedback gain.

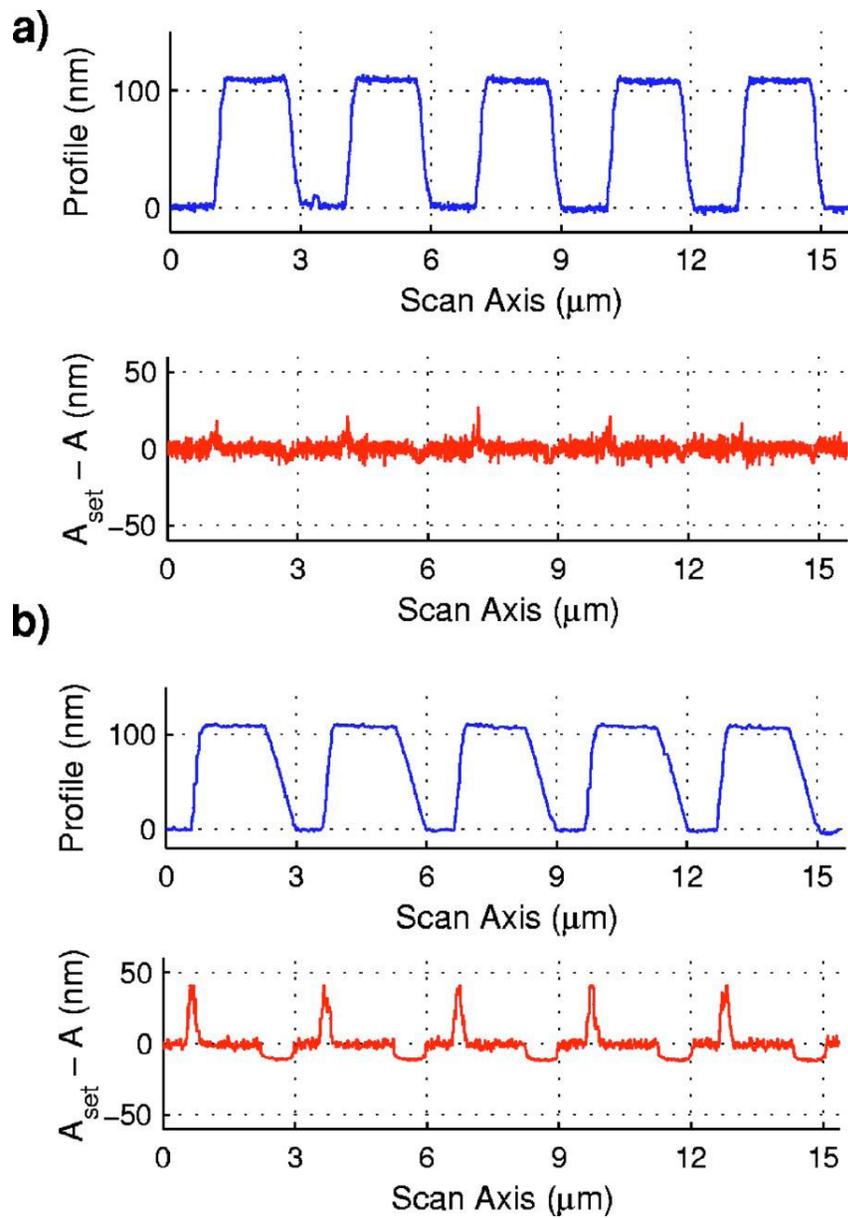

FIG. 6. The scan profile and the error signal ($A_{set} - A$) for the scan speeds of (a) X = 2 μm/s and (b) 5X when a standard PI controller is employed.

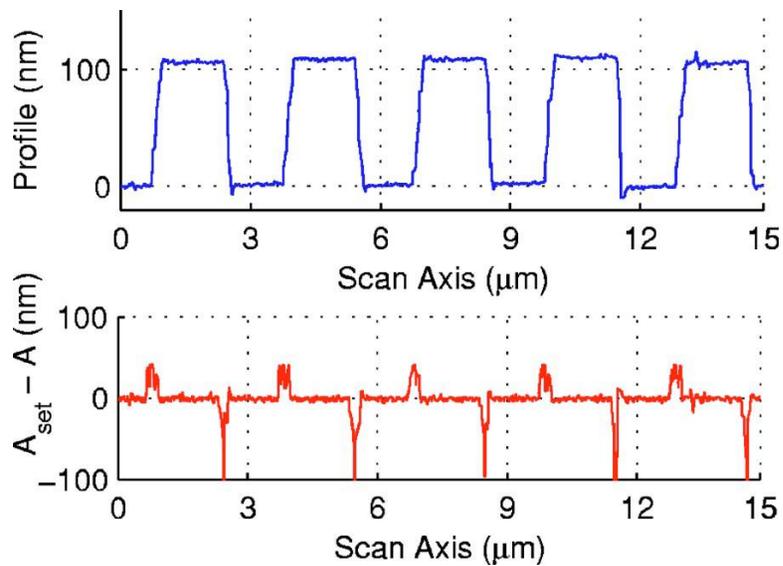

FIG. 7. The scan profile and the error signal for the scan speed of 5X when adaptive Q-controller is active.

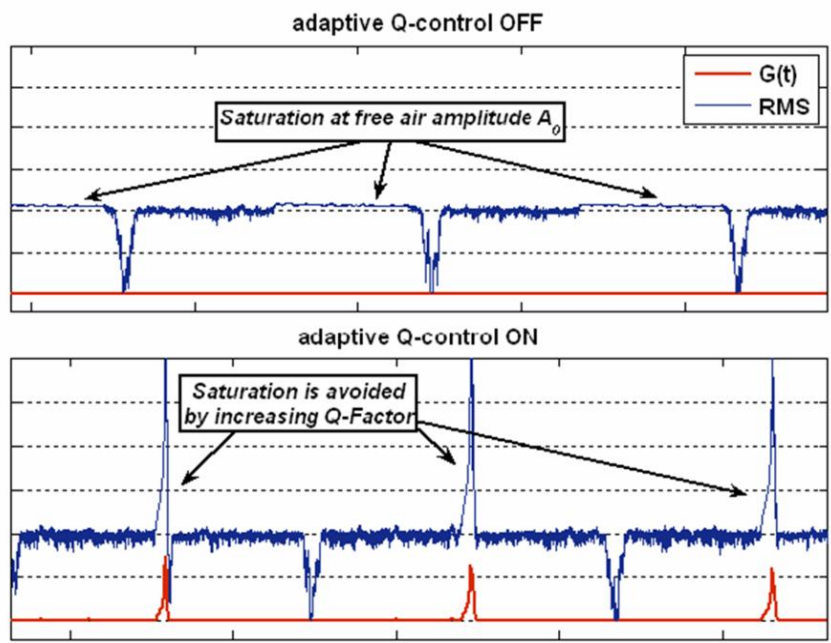

FIG. 8. When the adaptive Q-controller is active (b), the feedback gain G(t) is increased automatically such that the error saturation shown in (a) is avoided.

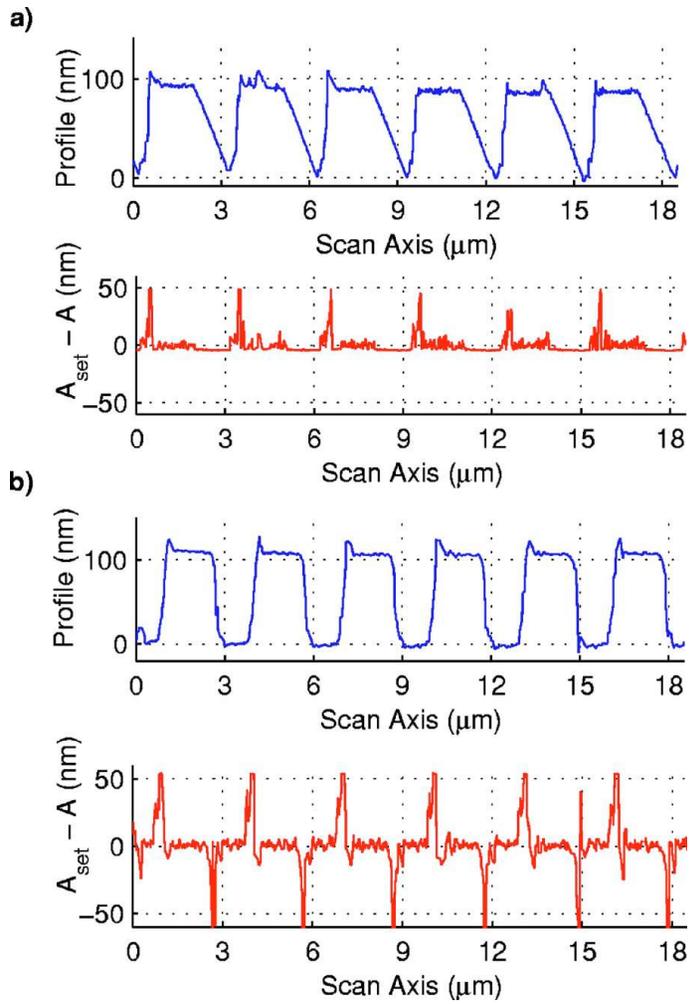

FIG. 9. The scan profile and the error signal for the scan speed of 10X when the standard Q-controller (a) and the adaptive Q-controller (b) are employed. Note that $Q_{set}$ is set to a value lower than its native one to increase the scan speed and $A_{set}$ is set to a value close to the free air amplitude $A_0$ to reduce the tapping forces simultaneously.